\documentclass[nofootinbib,preprint,floatfix,prb,tightenlines]{revtex4}
\usepackage{graphicx}
\usepackage{amsfonts}
\usepackage{amssymb}
\usepackage{amsmath}
\usepackage{latexsym}
\usepackage{subfloat}

\begin{document}

\preprint{gr-qc/0309006}

\title{Universes encircling 5-dimensional black holes}

\author{Sanjeev S.~Seahra}
\email{ssseahra@uwaterloo.ca} \affiliation{Department of Physics,
University of Waterloo, Waterloo, Ontario, N2L 3G1, Canada}

\author{Paul S.~Wesson}
\email{wesson@astro.uwaterloo.ca}%
\affiliation{Department of Physics, University of Waterloo,
Waterloo, Ontario, N2L 3G1, Canada}%

\date{September 1, 2003}

\setlength\arraycolsep{2pt}
\newcommand*{\di}{\partial}
\newcommand*{\V}{{\mathcal V}^{(k)}_d}
\newcommand*{\volume}{\sqrt{\sigma^{(k,d)}}}
\newcommand*{\OneTwo}{{(1,2)}}
\newcommand*{\onetwo}{{1,2}}
\newcommand*{\Lm}{{\mathcal L}_m}
\newcommand*{\stm}{{\textsc{stm}}}
\newcommand*{\Hm}{{\mathcal H}_m}
\newcommand*{\hatHm}{\hat{\mathcal H}_m}
\newcommand*{\Ldust}{{\mathcal L}_\mathrm{dust}}
\newcommand*{\maxsym}{{\mathbb S}_d^{(k)}}
\newcommand*{\ansatz}{{\emph{ansatz}}}
\newcommand*{\ds}[1]{ds^2_\text{\tiny{($#1$)}}}
\newcommand*{\kret}[1]{\mathfrak{K}_\text{\tiny{($#1$)}}}

\begin{abstract}

We clarify the status of two known solutions to the 5-dimensional
vacuum Einstein field equations derived by Liu, Mashhoon \& Wesson
(\textsc{lmw}) and Fukui, Seahra \& Wesson (\textsc{fsw}),
respectively. Both 5-metrics explicitly embed 4-dimensional
Friedman-Lema{\^\i}tre-Robertson-Walker cosmologies with a wide
range of characteristics.  We show that both metrics are also
equivalent to 5-dimensional topological black hole (\textsc{tbh})
solutions, which is demonstrated by finding explicit coordinate
transformations from the \textsc{tbh} to \textsc{lmw} and
\textsc{fsw} line elements.  We argue that the equivalence is a
direct consequence of Birkhoff's theorem generalized to 5
dimensions. Finally, for a special choice of parameters we plot
constant coordinate surfaces of the \textsc{lmw} patch in a
Penrose-Carter diagram.  This shows that the \textsc{lmw}
coordinates are regular across the black and/or white hole
horizons.

\end{abstract}

\maketitle

\section{Introduction}\label{sec:intro}

Over the past few years, there has been a marked resurgence of
interest in models with non-compact or large extra-dimensions.
Three examples of such scenarios immediately come to mind ---
namely the braneworld models of Randall \& Sundrum
\cite[henceforth \textsc{rs}]{Ran99a,Ran99b} and Arkani-Hamed,
Dimopoulos \& Dvali \cite[henceforth
\textsc{add}]{Ark98a,Ant98,Ark98b}, as well as the older
Space-Time-Matter (\textsc{stm}) theory \cite{Wes99}. The
\textsc{rs} model is motivated from certain ideas in string
theory, which suggest that the particles and fields of the
standard model are naturally confined to a lower-dimensional
hypersurface living in a non-compact, higher-dimensional bulk
manifold.  The driving goal behind the \textsc{add} picture is to
explain the discrepancy in scale between the observed strength of
the gravitational interaction and the other fundamental forces.
This is accomplished by noting that in generic higher-dimensional
models with compact extra dimensions, the bulk Newton's constant
is related to the effective 4-dimensional constant by factors
depending on the size and number of the extra dimensions. Finally,
\stm~or induced matter theory proposes that our universe is an
embedded 4-surface in a vacuum 5-manifold.  In this picture, what
we perceive to be the source in the 4-dimensional Einstein field
equations is really just an artifact of the embedding; or in other
words, conventional matter is induced from higher-dimensional
geometry.

Regardless of the motivation, if extra dimensions are to be taken
seriously then it is useful to have as many solutions of the
higher-dimensional Einstein equations at our disposal as possible.
These metrics serve as both arenas in which to test the
feasibility of extra dimensions, as well as guides as to where
4-dimensional general relativity may break down.  This simplest
type of higher-dimensional field equations that one might consider
is the 5-dimensional vacuum field equations $\hat{R}_{AB} = 0$.
(In this paper, uppercase Latin indices run $0 \ldots 4$ while
lowercase Greek indices run $0 \ldots 3$, and 5-dimensional
curvature tensors are distinguished from the 4-dimensional
counterparts by hats.  Also, commas in subscripts indicate partial
differentiation.) This condition is most relevant to the
\stm~scenario, but can also be applied to the \textsc{rs} or
\textsc{add} pictures. The are a fair number of known solutions
that embed 4-manifolds of cosmological or spherically-symmetric
character; one can consult the book by Wesson \cite{Wes99} for an
accounting of these metrics.

However, when searching for new solutions to vacuum field
equations, one must keep in mind a known peril from 4-dimensional
work; i.e., any new solution could be a previously discovered
metric written down in terms of strange coordinates.  Our purpose
in this paper is to demonstrate that two 5-dimensional vacuum
solutions in the literature are actually isometric to a
generalized 5-dimensional Schwarzschild manifold.  Both of these
solutions have been previously analyzed in the context of
4-dimensional cosmology because they both embed submanifolds with
line elements matching that of standard
Friedman-Lema{\^\i}tre-Robertson-Walker (\textsc{flrw}) models
with flat, spherical, or hyperbolic spatial sections.  In Section
\ref{sec:LMW}, we discuss the first of these 5-metrics, which was
originally written down by Liu \& Mashhoon \cite{Liu95} and later
rediscovered in a different form by Liu \& Wesson \cite{Liu01}. We
will see that this metric naturally embeds \textsc{flrw} models
with fairly general, but not unrestricted, scale factor behaviour.
Several different authors have considered this metric in a number
of different contexts \cite{Pon01,Liu02,Sea03,Wan03}, including
the \textsc{rs} braneworld scenario. The second 5-metric --- which
was discovered by Fukui, Seahra \& Wesson \cite{Fuk01} and is the
subject of Section \ref{sec:FSW}
--- also embeds \textsc{flrw} models with all types of spatial
curvature, but the scale factor is much more constrained.  We will
pay special attention to the characteristics of the embedded
cosmologies in each solution, as well as the coordinate invariant
geometric properties of the associated bulk manifolds.

The latter discussion will reveal that not only do the
Liu-Mashhoon-Wesson (\textsc{lmw}) and Fukui-Seahra-Wesson
(\textsc{fsw}) metrics have a lot in common with one another, they
also exhibit many properties similar to that of the topological
black hole (\textsc{tbh}) solution of the 5-dimensional vacuum
field equations, which we introduce in Section
\ref{sec:topological}. This prompts us to suspect that the
\textsc{lmw} and \textsc{fsw} solutions are actually isometric to
topological black hole manifolds.  We confirm this explicitly by
finding transformations from standard black hole to \textsc{lmw}
and \textsc{fsw} coordinates in Sections \ref{sec:coord LMW} and
\ref{sec:coord FSW} respectively.  We argue that the equivalence
of the three metrics is actually a consequence of a
higher-dimensional version of Birkhoff's theorem in Section
\ref{sec:birkhoff}.  In Section \ref{sec:coord Kruskal}, we
discuss which portion of the extended 5-dimensional Kruskal
manifold is covered by the \textsc{lmw} coordinate patch and
obtain Penrose-Carter embedding diagrams for a particular case.
Section \ref{sec:summary} summarizes and discusses our results.

\section[Two 5-metrics with \textsc{flrw} submanifolds]{Two 5-metrics with
FLRW submanifolds}

In this section, we introduce two 5-metrics that embed
4-dimensional \textsc{flrw} models.  Both of these are solutions
of the 5-dimensional vacuum field equations, and are hence
suitable manifolds for \textsc{stm} theory.
 Our goals are to illustrate what subset of all possible \textsc{flrw}
models can be realized as hypersurfaces contained within these
manifolds, and to find out about any 5-dimensional curvature
singularities or geometric features that may be present.

\subsection{The Liu-Mashhoon-Wesson metric}
\label{sec:LMW}

Consider a 5-dimensional manifold
$(M_\text{\textsc{lmw}},g_{AB})$. We define the \textsc{lmw}
metric \ansatz~as:
\begin{equation}\label{LMW metric}
    ds^2_\text{\textsc{lmw}} = \frac{ a^2_{,t}(t,\ell) }{\mu^2(t)}
    dt^2 - a^2(t,\ell) \, d\sigma_{(k,3)}^2 - d\ell^2.
\end{equation}
Here, $a(t,\ell)$ and $\mu(t)$ are undetermined functions, and
$d\sigma_{(k,3)}^2$ is the line element on maximally symmetric
3-spaces $\mathbb{S}_3^{(k)}$ with curvature index $k = +1,0,-1$:

\begin{equation}
    d\sigma_{(k,3)}^2 = d\psi^2 + S_k^2(\psi) (d\theta^2 +
    \sin^2\theta\,d\varphi^2),
\end{equation}
where
\begin{equation}\label{S_k def}
    S_k(\psi) \equiv
    \begin{cases}
        \sin\psi, & k = +1, \\
        \psi, & k = 0, \\
        \sinh\psi, & k = -1,
    \end{cases}
\end{equation}
It is immediately obvious that the $\ell =$ constant hypersurfaces
$\Sigma_\ell$ associated with (\ref{LMW metric}) have the
structure of \textsc{flrw} models: $\mathbb{R} \times
\mathbb{S}_3^{(k)}$.  We should note that the original papers
(refs.~\onlinecite{Liu95} and \onlinecite{Liu01}) did not really
begin with a metric \ansatz~like (\ref{LMW metric}); rather, the
$g_{tt}$ component of the metric was initially taken to be some
general function of $t$ and $\ell$. But one rapidly closes in on
the above line element by direct integration of one component of
the vacuum field equations $\hat{R}_{AB} = 0$; namely,
$\hat{R}_{t\ell} = 0$. The other components are satisfied if
\begin{equation}\label{a soln}
    a^2(t,\ell) = [ \mu^2(t) + k ] \ell^2 + 2 \nu(t) \ell + \frac{
    \nu^2(t) + \mathcal{K} }{ \mu^2(t) + k },
\end{equation}
where $\mathcal{K}$ is an integration constant.  As far as the
field equations are concerned, $\mu(t)$ and $\nu(t)$ are
\emph{completely arbitrary} functions of time.  However, we should
constrain them by appending the condition
\begin{equation}
    a(t,\ell) \in \mathbb{R}^+ \quad \Rightarrow \quad a^2(t,\ell) > 0
\end{equation}
to the system.  This restriction ensures that the metric signature
is $(+----)$ and $t$ is the only timelike coordinate.  Now, if $a$
is taken to be real, then it follows that $\nu$ must be real as
well.  Regarding (\ref{a soln}) as a quadratic equation in $\nu$,
we find that there are real solutions only if the quadratic
discriminant is non-negative.  This condition translates into
\begin{equation}\label{inequality}
    \mathcal{K} \le a^2(t,\ell) [\mu^2(t) + k].
\end{equation}
If $\mathcal{K}$ is positive this inequality implies that we must
choose $\mu(t)$ such that $\mu^2 + k > 0$. This relation will be
important shortly.

The reason that this solution is of interest is that the induced
metric on $\ell =$ constant hypersurfaces is isometric to the
standard \textsc{flrw} line element.  To see this explicitly,
consider the line element on the $\ell = \ell_0$ 4-surface:
\begin{equation}\label{original 4-metric}
    \ds{\Sigma_\ell} = \frac{a_{,t}^2(t,\ell_0)}{\mu^2(t)} dt^2 -
    a^2(t,\ell_0) \, d\sigma_{(k,3)}^2.
\end{equation}
Let us perform the 4-dimensional coordinate transformation
\begin{equation}\label{4-transform}
    \Theta(t) = \int_t \frac{a_{,u}(u,\ell_0)}{\mu(u)} du \quad
    \Rightarrow \quad \mu(t(\Theta)) = {\mathcal A}'(\Theta),
\end{equation}
where
\begin{equation}
    {\mathcal{A}}(\Theta) = a(t(\Theta),\ell_0),
\end{equation}
and we use a prime to denote the derivative of functions of a
single argument.  This puts the induced metric in the
\textsc{flrw} form
\begin{equation}\label{altered 4-metric}
    \ds{\Sigma_\ell} = d\Theta^2 -
    {\mathcal{A}}^2(\Theta) \, d\sigma_{(k,3)}^2,
\end{equation}
where $\Theta$ is the cosmic time and ${\mathcal A}(\Theta)$ is
the scale factor.

So, the geometry of each of the $\Sigma_\ell$ hypersurfaces is
indeed of the \textsc{flrw}-type.  But what kind of cosmologies
can be thus embedded? Well, if we rewrite the inequality
(\ref{inequality}) in terms of $\mathcal{A}$ and ${\mathcal A}'$
we obtain
\begin{equation}\label{inequal 1}
    \mathcal{K} \le {\mathcal A}^2 ( {\mathcal A}'^2 + k ).
\end{equation}
Since $\mathcal A$ is to be interpreted as the scale factor of
some cosmological model, it satisfies the Friedman equation:
\begin{equation}
    {\mathcal A}'^2 - \tfrac{1}{3} \kappa_4^2 \rho {\mathcal A}^2 = -k.
\end{equation}
Here, $\rho$ is the total density of the matter-energy in the
cosmological model characterized by ${\mathcal A}(\Theta)$ and
$\kappa_4^2 = 8\pi G$ is the usual coupling constant in the
4-dimensional Einstein equations. This implies a relation between
the density of the embedded cosmologies and the choice of $\mu$:
\begin{equation}\label{mu density reln}
    \mu^2 + k = \tfrac{1}{3} \kappa_4^2 \rho {\mathcal A}^2.
\end{equation}
This into the inequality (\ref{inequal 1}) yields
\begin{equation}\label{inequality 2}
    \mathcal{K} \le \tfrac{1}{3} \kappa_4^2 \rho {\mathcal A}^4.
\end{equation}
Therefore, we can successfully embed a given \textsc{flrw} model
on a $\Sigma_\ell$ 4-surface in the \textsc{lmw} solution if the
total density of the model's cosmological fluid and scale factor
satisfy (\ref{inequality 2}) for all $\Theta$.  An obvious
corollary of this is that we can embed any \textsc{flrw} model
with $\rho > 0$ if $\mathcal{K} < 0$.

There is one other point about the intrinsic geometry of the
$\Sigma_\ell$ hypersurfaces that needs to be made.  Notice that
our 4-dimension coordinate transformation (\ref{4-transform}) has
\begin{equation}
    \frac{d\Theta}{dt} = \frac{a_{,t}}{\mu},
\end{equation}
which means that the associated Jacobian vanishes whenever $a_{,t}
= 0$.  Therefore, the transformation is really only valid in
between the turning points of $a$.  Also notice that the original
4-metric (\ref{original 4-metric}) is badly behaved when $a_{,t}
=0$, but the transformed one (\ref{altered 4-metric}) is not when
$\mathcal{A}' = 0$.  We can confirm via direct calculation that
the Ricci scalar for (\ref{original 4-metric}) is
\begin{equation}
    {}^{(4)}R = - \frac{6\mu}{a} \frac{d\mu}{dt} \left( \frac{\di
    a}{\di t} \right)^{-1} - \frac{6}{a^2} (\mu^2 +k).
\end{equation}
We see that ${}^{(4)}R$ diverges when $a_{,t} = 0$, provided that
$\mu \mu_{,t} /a \ne 0$.  Therefore, there can be genuine
curvature singularities in the intrinsic 4-geometry at the turning
points of $a$.  These features are hidden in the altered line
element (\ref{altered 4-metric}) because the coordinate
transformation (\ref{4-transform}) is not valid in the immediate
vicinity of any singularities, hence the $\Theta$-patch cannot
cover those regions (if they exist).  We mention that this
4-dimensional singularity in the \textsc{lmw} metric has been
recently investigated by Xu, Liu and Wang \cite{Xu03}, who have
interpreted it as a 4-dimensional event horizon.

Now, let us turn our attention to some of the 5-dimensional
geometric properties of $M_\text{\textsc{lmw}}$.  We can test for
curvature singularities in this 5-manifold by calculating the
Kretschmann scalar:
\begin{equation}\label{K LMW}
    \mathfrak{K}_\text{\textsc{lmw}} \equiv \hat{R}^{ABCD} \hat{R}_{ABCD} =
    \frac{72 \mathcal{K}^2}{a^8(t,\ell)}.
\end{equation}
We see there is a singularity in the 5-geometry along the
hypersurface $a(t,\ell) = 0$.  (Of course, whether or not
$a(t,\ell) = 0$ for any $(t,\ell) \in \mathbb{R}^2$ depends on the
choice of $\mu$ and $\nu$.)  This singularity is essentially a
line-like object because the radius $a$ of the 3-dimensional
$\mathbb{S}_3^{(k)}$ subspace vanishes there.  Other tools for
probing the 5-geometry are Killing vector fields on
$M_\text{\textsc{lmw}}$.  Now, there are by definition 6 Killing
vectors associated with symmetry operations on
$\mathbb{S}_3^{(k)}$, but there is also at least one Killing
vector that is orthogonal to that submanifold. This vector field
is given by
\begin{equation}
    \xi^\text{\textsc{lmw}}_A dx^A = \frac{ a_{,t} }{\mu} \sqrt{ h(a) + \mu^2(t) } \, dt +
    \sqrt{ a_{,\ell}^2 - h(a) } \, d\ell.
\end{equation}
Here, we have defined
\begin{equation}
    h(x) \equiv k - \frac{\mathcal{K}}{x^2}.
\end{equation}
Using the explicit form of $a(t,\ell)$ from equation (\ref{a
soln}), we can verify that $\xi$ satisfies Killing's equation
\begin{equation}\label{Killing equation}
    \nabla_B \xi_A^\text{\textsc{lmw}} + \nabla_A \xi_B^\text{\textsc{lmw}} = 0,
\end{equation}
via computer.  Also using (\ref{a soln}), we can calculate the
norm of $\xi^\text{\textsc{lmw}}$, which is given by
\begin{equation}\label{killing LMW}
    \xi^\text{\textsc{lmw}} \cdot \xi^\text{\textsc{lmw}} = h(a).
\end{equation}
This vanishes at $ka^2 = \mathcal{K}$.  So, if $k\mathcal{K}>0$
the 5-manifold contains a Killing horizon.  If the horizon exists
then $\xi^\text{\textsc{lmw}}$ will be timelike for $|a| >
\sqrt{|\mathcal{K}|}$ and spacelike for $|a| <
\sqrt{|\mathcal{K}|}$.

To summarize, we have seen that \textsc{flrw} models satisfying
(\ref{inequality 2}) can be embedded on a $\Sigma_\ell$ 4-surface
within the \textsc{lmw} metric, but that there are 4-dimensional
curvature singularities wherever $a_{,t} = 0$.  The \textsc{lmw}
5-geometry also possesses a line-like singularity where $a(t,\ell)
= 0$, as well as a Killing horizon across which the norm of
$\xi^\text{\textsc{lmw}}$ changes sign.

\subsection{The Fukui-Seahra-Wesson metric}
\label{sec:FSW}

For the time being, let us set aside the \textsc{lmw} metric and
concentrate on the \textsc{fsw} solution.  On a certain 5-manifold
$(M_\text{\textsc{fsw}},g_{AB})$, this is given by the line
element
\begin{subequations}\label{FSW}
\begin{eqnarray}\label{FSW metric}
    ds^2_\text{\textsc{fsw}} & = & d\tau^2 - {b}^2(\tau,w) \, d\sigma_{(k,3)}^2 - \frac{
    b^2_{,w}(\tau,w)}{ \zeta^2(w)} dw^2, \\ \label{FSW scale}
    {b}^2(\tau,w)& = & [\zeta^2(w) - k] \tau^2 + 2\chi(w) \tau + \frac{
    \chi^2(w) - \mathcal{K} }{ \zeta^2(w) - k }.
\end{eqnarray}
\end{subequations}
This metric (\ref{FSW metric}) is a solution of the 5-dimensional
vacuum field equations $\hat{R}_{AB} = 0$ with $\zeta(w)$ and
$\chi(w)$ as arbitrary functions.  Just as before, we call
equation (\ref{FSW metric}) the \textsc{fsw} metric \ansatz, even
though it was not the technical starting point of the original
paper\cite{Fuk01}.  We have written (\ref{FSW}) in a form somewhat
different from that of ref.~\onlinecite{Fuk01}; to make contact
with their notation we need to make the correspondences
\begin{subequations}
\begin{eqnarray}
    \left[F(w)\right]_{\text{\textsc{fsw}}} & \equiv & k - \zeta^2(w), \\
    \left[h(w)\right]_{\text{\textsc{fsw}}} & \equiv & [\chi^2(w) + \mathcal{K} ] / [ \zeta^2(w) - k ], \\
    \left[g(w)\right]_{\text{\textsc{fsw}}} & \equiv & 2 \chi(w), \\
    \left[\mathcal{K}\right]_{\text{\textsc{fsw}}} & \equiv &
    -4\mathcal{K},
\end{eqnarray}
\end{subequations}
where $\left[\cdots\right]_\text{\textsc{fsw}}$ indicates a
quantity from the original \textsc{fsw} work.  A cursory
comparison between the \textsc{lmw} and \textsc{fsw} vacuum
solutions reveals that both metrics have a similar structure,
which prompts us to wonder about any sort of fundamental
connection between them. We defer this issue to the next section,
and presently concern ourselves with the properties of the
\textsc{fsw} solution in its own right.

Just as for the \textsc{lmw} metric, we can identify hypersurfaces
in the \textsc{fsw} solution with \textsc{flrw} models.
Specifically, the induced metric on $w = w_0$ hypersurfaces
$\Sigma_w$ is
\begin{equation}
    \ds{\Sigma_w} = d\tau^2 - b^2(\tau,w_0) \, d\sigma_{(k,3)}^2.
\end{equation}
We see that for the universes on $\Sigma_w$, $\tau$ is the cosmic
time and $b(\tau,w_0)$ is the scale factor.  It is useful to
perform the following linear transformation on $\tau$:
\begin{equation}
    \tau(\Theta) = \Theta - \frac{\chi_0}{\zeta_0^2 - k},
\end{equation}
where we have defined $\zeta_0 \equiv \zeta(w_0)$ and $\chi_0
\equiv \chi(w_0)$.  This puts the induced metric into the form
\begin{subequations}
\begin{eqnarray}
    \ds{\Sigma_w} & = & d\Theta^2 -
    {\mathcal B}^2(\Theta) \, d\sigma_{(k,3)}^2, \\
    \label{transformed FSW scale}
    {\mathcal B}(\Theta) & = & \sqrt{\frac{ (\zeta_0^2 - k)^2 \Theta^2
    - \mathcal{K} }{ \zeta_0^2 - k }}.
\end{eqnarray}
\end{subequations}
Unlike the \textsc{lmw} case, the cosmology on the $\Sigma_w$
hypersurfaces has restrictive properties.  If $\zeta_0^2 - k > 0$,
the scale factor ${\mathcal B}(\Theta)$ has the shape of one arm
of a hyperbola with a semi-major axis of length
$\sqrt{-\mathcal{K}/(\zeta_0^2 - k)}$.  Note that this length may
be complex depending on the values of $\zeta_0$, $k$ and
$\mathcal{K}$.  That is, the scale factor may not be defined for
all $\Theta \in \mathbb{R}$.  When this is the case, the embedded
cosmologies involve a big bang and/or a big crunch. Conversely, it
is not hard to see if $\zeta_0^2 - k < 0$ and $\mathcal{K} > 0$
then the cosmology is re-collapsing; i.e., there is a big bang and
a big crunch.  However, if $\zeta_0^2 - k < 0$ and $\mathcal{K}
\le 0$, then there is no $\Theta$ interval where the scale factor
is real. We have summarized the basic properties of the embedded
cosmologies in Table \ref{tab:properties}. Finally, we note that
if $\zeta_0^2 - k > 0$ then
\begin{equation}
    \lim_{\Theta \rightarrow \infty} \mathcal{B}(\Theta) = (\zeta_0^2 -
    k)^{1/2} \Theta.
\end{equation}
Hence, the late time behaviour of such models approaches that of
the empty Milne universe.
\begin{table}
\begin{centering}
\begin{tabular}{|c|cc|} \hline
{}&$\zeta_0^2 - k > 0$ & $\zeta_0^2 - k < 0 $\\ \hline\hline
$\mathcal{K} > 0$ & big bang & big bang and big crunch \\
$\mathcal{K} = 0$ & big bang & $\mathcal{B} \in \mathbb{C}$ for all $\Theta \in \mathbb{R}$ \\
$\mathcal{K} < 0$ & no big bang/crunch & $\mathcal{B} \in \mathbb{C}$ for all $\Theta \in \mathbb{R}$ \\
\hline
\end{tabular}
\caption[Types of cosmologies embedded in the \textsc{fsw}
metric]{Characteristics of the 4-dimensional cosmologies embedded
on the $\Sigma_w$ hypersurfaces in the \textsc{fsw}
metric}\label{tab:properties}
\end{centering}
\end{table}

Lake \cite{Lak01} has calculated the Kretschmann scalar for vacuum
5-metrics of the \textsc{fsw} type.  When his formula is applied
to (\ref{FSW}), we obtain:
\begin{equation}\label{K FSW}
    \mathfrak{K}_\text{\textsc{fsw}} \equiv \hat{R}^{ABCD} \hat{R}_{ABCD} =
    \frac{72 \mathcal{K}^2}{b^8(\tau,w)}.
\end{equation}
As for the \textsc{lmw} manifold, this implies the existence of a
line-like singularity in the 5-geometry at $b(\tau,w) = 0$.  We
also find that there is a Killing vector on
$M_\text{\textsc{fsw}}$, which is given by
\begin{subequations}
\begin{eqnarray}
    \xi^\text{\textsc{fsw}}_A dx^A & = & \sqrt{b_{,\tau} + h(b)} \, d\tau + \frac{
    b_{,w}}{\zeta} \sqrt{\zeta^2 - h(b)} \, dw \\
    0 & = & \nabla_A \xi_B^\text{\textsc{fsw}} + \nabla_B \xi_A^\text{\textsc{fsw}}.
\end{eqnarray}
\end{subequations}
The norm of this Killing vector is relatively easily found by
computer:
\begin{equation}\label{killing FSW}
    \xi^\text{\textsc{fsw}} \cdot \xi^\text{\textsc{fsw}} = h(b).
\end{equation}
Hence, there is a Killing horizon in $M_\text{\textsc{fsw}}$ where
$h(b) = 0$. Obviously, the $\xi^\text{\textsc{fsw}}$ Killing
vector changes from timelike to spacelike --- or \emph{vice versa}
--- as the horizon is traversed.

In summary, we have seen how \textsc{flrw} models with scale
factors of the type (\ref{FSW scale}) are embedded in the
\textsc{fsw} solution.  We found that there is a line-like
curvature singularity in $M_\text{\textsc{fsw}}$ at $b(\tau,w) =
0$ and the bulk manifold has a Killing horizon where the magnitude
of $\xi^\text{\textsc{fsw}}$ vanishes.

\section[Connection to the 5D topological black hole manifold]
{Connection to the 5-dimensional topological black hole manifold}
\label{sec:topological}

When comparing equations (\ref{K LMW}) and (\ref{K FSW}), or
(\ref{killing LMW}) and (\ref{killing FSW}), it is hard not to
believe that there is some sort of fundamental connection between
the \textsc{lmw} and \textsc{fsw} metrics. We see that
\begin{equation}
    \mathfrak{K}_\text{\textsc{lmw}} = \mathfrak{K}_\text{\textsc{fsw}}, \quad \xi^\text{\textsc{lmw}}
    \cdot \xi^\text{\textsc{lmw}} = \xi^\text{\textsc{fsw}} \cdot
    \xi^\text{\textsc{fsw}},
\end{equation}
if we identify $a(t,\ell) = b(\tau,w)$.  Also, we notice that the
\textsc{lmw} solution can be converted into the \textsc{fsw}
metric by the following set of transformations/Wick rotations
\cite{note}:
\begin{equation}
\begin{array}{rclcrcl}
    \psi & \rightarrow & i\psi, & \quad &
    t & \rightarrow & w, \\ \ell & \rightarrow & \tau, & &
    k & \rightarrow & -k, \\
    \mathcal{K} & \rightarrow & -\mathcal{K}, & & ds_\text{\textsc{lmw}} & \rightarrow &
    i\,ds_\text{\textsc{fsw}}.
\end{array}
\end{equation}
These facts lead us to the strong suspicion that the \textsc{lmw}
and \textsc{fsw} metrics actually describe the same 5-manifold.

But which 5-manifold might this be?  We established in the
previous section that both the \textsc{lmw} and \textsc{fsw}
metrics involve a 5-dimensional line-like curvature singularity
and Killing horizon if $k\mathcal{K}>0$. This reminds us of
another familiar manifold: that of a black hole.  Consider the
metric of a ``topological'' black hole (\textsc{tbh})  on a
5-manifold $(M_\text{\textsc{tbh}},g_{AB})$:
\begin{equation}\label{BH metric}
    ds^2_\text{\textsc{tbh}} = h(R) \, dT^2 - h^{-1}(R) \, dR^2 - R^2 \, d\sigma_{(k,3)}^2.
\end{equation}
The adjective ``topological'' comes from the fact that the
manifold has the structure $\mathbb{R}^2 \times
\mathbb{S}_3^{(k)}$, as opposed to the familiar $\mathbb{R}^2
\times S_3$ structure commonly associated with spherical symmetry
in 5-dimensions.  That is, the surfaces $T = $ constant and $R =$
constant are not necessarily 3-spheres for the topological black
hole; it is possible that they have flat or hyperbolic geometry.
One can confirm by direct calculation that (\ref{BH metric}) is a
solution of $\hat{R}_{AB} = 0$ for any value of $k$, and that the
constant $\mathcal{K}$ that appears in $h(R)$ is related to the
mass of the central object.  The Kretschmann scalar on
$M_\text{\textsc{tbh}}$ is
\begin{equation}\label{K TBH}
    \mathfrak{K}_\text{\textsc{tbh}} = \hat{R}^{ABCD} \hat{R}_{ABCD} = \frac{72
    \mathcal{K}^2}{R^8},
\end{equation}
implying a line-like curvature singularity at $R = 0$. There is an
obvious Killing vector in this manifold, given by
\begin{equation}
    \xi^\text{\textsc{tbh}}_A dx^A = h(R) \, dT.
\end{equation}
The norm of this vector is trivially
\begin{equation}\label{killing TBH}
    \xi^\text{\textsc{tbh}} \cdot \xi^\text{\textsc{tbh}} = h(R).
\end{equation}
There is therefore a Killing horizon in this space located at
$kR^2 = \mathcal{K}$.

Now, equations (\ref{K TBH}) and (\ref{killing TBH}) closely match
their counterparts for the \textsc{lmw} and \textsc{fsw} metrics,
which inspires the hypothesis that not only are the \textsc{lmw}
and \textsc{fsw} isometric to one another, they are also isometric
to the metric describing topological black holes. However, while
these coincidences provide fairly compelling circumstantial
evidence that the \textsc{lmw}, \textsc{fsw}, and \textsc{tbh}
metrics are equivalent, we do not have conclusive proof --- that
will come in the next section.

\section{Coordinate transformations}\label{sec:transformations}

In this section, our goal is to prove the conjecture that the
\textsc{lmw}, \textsc{fsw}, and \textsc{tbh} solutions and the
5-dimensional vacuum field equations are isometric to one another.
We will do so by finding two explicit coordinate transformations
that convert the \textsc{tbh} metric to the \textsc{lmw} and
\textsc{fsw} metrics respectively.  This is sufficient to prove
the equality of all three solutions, since it implies that one can
transform from the \textsc{lmw} to the \textsc{fsw} metric
--- or \emph{vice versa} --- via a two-stage procedure.

\subsection[Schwarzschild to Liu-Mashhoon-Wesson coordinates]
{Transformation from Schwarzschild to Liu-Mashhoon-Wesson
coordinates} \label{sec:coord LMW}

We first search for a coordinate transformation that takes the
\textsc{tbh} line element (\ref{BH metric}) to the \textsc{lmw}
line element (\ref{LMW metric}). We take this transformation to be
\begin{equation}\label{transform}
    R = {\mathcal R}(t,\ell), \quad T = {\mathcal T}(t,\ell).
\end{equation}
Notice that we have \emph{not} assumed $R = a(t,\ell)$
--- as may have been expected from the discussion of the previous section ---
in order to stress that we are starting with a general coordinate
transformation.  We will soon see that by demanding that this
transformation forces the \textsc{tbh} metric into the form of the
\textsc{lmw} metric \ansatz, we can recover $R = a(t,\ell)$ with
$a(t,\ell)$ given explicitly by (\ref{a soln}).  In other words,
the coordinate transformation specified in this section will fix
the functional form of $a(t,\ell)$ in a manner independent of the
direct attack on the vacuum field equations found in
refs.~\onlinecite{Liu95} and \onlinecite{Liu01}.

When (\ref{transform}) is substituted into (\ref{BH metric}), we
get
\begin{eqnarray}\nonumber
    ds^2_\text{\textsc{tbh}} & = & \left[ h({\mathcal R}) {\mathcal T}_{,t}^2 - \frac{ {\mathcal R}_{,t}^2 }{
    h({\mathcal R}) }  \right] dt^2 + 2 \left[ h({\mathcal R}) {\mathcal T}_{,t} {\mathcal T}_{,\ell} -
    \frac{ {\mathcal R}_{,t} {\mathcal R}_{,\ell} }{h({\mathcal R})} \right] dt\,d\ell +
    \\ & & \left[ h({\mathcal R})
    \mathcal{T}_{,\ell}^2 - \frac{ {\mathcal R}_{,\ell}^2 }{
    h({\mathcal R}) }  \right] d\ell^2 - {\mathcal R}^2(t,\ell) \, d\sigma_{(k,3)}^2.
\end{eqnarray}
For this to match equation (\ref{LMW metric}) with ${\mathcal
R}(t,\ell)$ instead of $a(t,\ell)$ we must have
\begin{subequations}\label{the transformation}
\begin{eqnarray}\label{one}
    \frac{{\mathcal R}_{,t}^2}{\mu^2(t)} & = & h({\mathcal R}) {\mathcal T}_{,t}^2 - \frac{ {\mathcal R}_{,t}^2 }{
    h({\mathcal R}) }, \\ \label{two} 0 & = & h({\mathcal R}) {\mathcal T}_{,t} {\mathcal T}_{,\ell} -
    \frac{ {\mathcal R}_{,t} {\mathcal R}_{,\ell} }{h({\mathcal R})}, \\ \label{three} -1 & = & h({\mathcal R})
    \mathcal{T}_{,\ell}^2 - \frac{ {\mathcal R}_{,\ell}^2 }{
    h({\mathcal R}) },
\end{eqnarray}
\end{subequations}
with $\mu(t)$ arbitrary.  Under these conditions, we find
\begin{equation}
    ds^2_\text{\textsc{tbh}} = \frac{ \mathcal{R}^2_{,t}(t,\ell) }{\mu^2(t)}
    dt^2 - \mathcal{R}^2(t,\ell) \, d\sigma_{(k,3)}^2 - dy^2,
\end{equation}
which is obviously the same as the \textsc{lmw} metric
\ansatz~(\ref{LMW metric}). However, the precise functional form
of $\mathcal{R}(t,\ell)$ has yet to be specified.

To solve for $\mathcal{R}(t,\ell)$, we note equations (\ref{one})
and (\ref{three}) can be rearranged to give
\begin{subequations}\label{T eqns}
\begin{eqnarray}
    \mathcal{T}_{,t} & = & \epsilon_t \frac{{\mathcal R}_{,t}}{h({\mathcal R})} \sqrt{ 1 +
    \frac{ h({\mathcal R}) }{\mu^2(t)}}, \\ \mathcal{T}_{,\ell} & = &
    \epsilon_\ell \frac{1}{h({\mathcal R})} \sqrt{ {\mathcal R}^2_{,\ell} -
    h({\mathcal R}) },
\end{eqnarray}
\end{subequations}
where $\epsilon_t = \pm 1$ and $\epsilon_\ell = \pm 1$.  Using
these in (\ref{two}) yields
\begin{equation}\label{R eqn}
    {\mathcal R}_{,\ell} = \pm \sqrt{h({\mathcal R}) + \mu^2(t)}.
\end{equation}
Our task it to solve the system of \textsc{pde}s formed by
equations (\ref{T eqns}) and (\ref{R eqn}) for ${\mathcal
T}(t,\ell)$ and ${\mathcal R}(t,\ell)$. Once we have accomplished
this, the coordinate transformation from (\ref{LMW metric}) to
(\ref{BH metric}) is found.

Using the definition of $h({\mathcal R})$, we can expand equation
(\ref{R eqn}) to get
\begin{equation}\label{ODE}
    \pm 1 = \frac{{\mathcal R}}{\sqrt{(\mu^2 + k){\mathcal R}^2 - \mathcal{K}}} \frac{\di {\mathcal R}}{
    \di \ell}.
\end{equation}
Integrating both sides with respect to $\ell$ yields
\begin{equation}
    \sqrt{(\mu^2 + k){\mathcal R}^2 - \mathcal{K}} = (\mu^2 + k)(\pm \ell + \gamma),
\end{equation}
where $\gamma = \gamma(t)$ is an arbitrary function of time.
Solving for ${\mathcal R}$ gives
\begin{equation}\label{R soln}
    R^2 = {\mathcal R}^2(t,\ell) = [ \mu^2(t) + k ] \ell^2 + 2 \nu(t) \ell + \frac{
    \nu^2(t) + \mathcal{K} }{ \mu^2(t) + k },
\end{equation}
where we have defined
\begin{equation}
    \nu(t) = \pm \gamma(t)[\mu^2(t) + k],
\end{equation}
which can be thought of as just an another arbitrary function of
time.  We have hence see that the functional form of
$\mathcal{R}(t,\ell)$ matches exactly the functional form of
$a(t,\ell)$ in equation (\ref{a soln}).  This is despite the fact
that the two expressions were derived by different means: (\ref{R
soln}) from conditions placed on a coordinate transformation, and
(\ref{a soln}) from the direct solution of the 5-dimensional
vacuum field equations.

When our solution for ${\mathcal R}(t,\ell)$ is put into equations
(\ref{T eqns}), we obtain a pair of \textsc{pde}s that expresses
the gradient of ${\mathcal T}$ in the $(t,\ell)$-plane as known
functions of the coordinates. This is analogous to a problem where
one is presented with the components of a 2-dimensional force and
is asked to find the associated potential.  The condition for
integrablity of the system is that the curl of the force vanishes,
which in our case reads
\begin{equation}
    0 \stackrel{\scriptstyle ?}{=} \epsilon_t \frac{\di}{\di \ell}
    \left( \frac{{\mathcal R}_{,t}}{h({\mathcal R})} \sqrt{ 1 +
    \frac{ h({\mathcal R}) }{\mu^2(t)}} \right) - \epsilon_\ell \frac{\di}{\di t}
    \left( \frac{1}{h({\mathcal R})}
    \sqrt{ {\mathcal R}_{,\ell}^2 - h({\mathcal R}) } \right).
\end{equation}
We have confirmed via computer that this condition holds when
${\mathcal R}(t,\ell)$ is given by equation (\ref{R soln}),
provide we choose $\epsilon_t = \epsilon_\ell = \pm 1$.  Without
loss of generality, we can set $\epsilon_t = \epsilon_\ell = 1$.
Hence, equations (\ref{T eqns}) are indeed solvable for ${\mathcal
T}(t,\ell)$ and a \emph{coordinate transformation from (\ref{BH
metric}) to (\ref{LMW metric}) exists}.

The only thing left is the tedious task of determining the
explicit form of ${\mathcal T}(t,\ell)$.  We spare the reader the
details and just quote the solution, which can be checked by
explicit substitution into (\ref{T eqns}). For $k = \pm 1$, we get
\begin{subequations}\label{T soln curved}
\begin{eqnarray}\nonumber
    {\mathcal T}(t,\ell) & = & \frac{1}{k} \int_t \left\{ \frac{1}{\mu(u)} \frac{d}{du}\nu(u)-
    \left[ \frac{\nu(u)}{\mu^2(u) + k} \right] \frac{d}{du} \mu(u) \right\}
    du +
    \\ & & \frac{1}{k} \left( \mu(t) \ell -
    \frac{\mathcal{K}}{2\sqrt{k\mathcal{K}}} \, \ln
    \frac{1 + \mathcal{X}(t,\ell)}{1-\mathcal{X}(t,\ell)} \right), \\ \mathcal{X}(t,\ell) & \equiv & \frac{k}{\sqrt{k\mathcal{K}}}
    \frac{[\mu^2(t) +k]\ell + \nu(t)}{\mu(t)}.
\end{eqnarray}
\end{subequations}
For $k = 0$, we obtain
\begin{eqnarray}\nonumber
    {\mathcal T}(t,\ell) & = &
    \frac{1}{\mathcal{K}} \int_t  \left\{ \frac{\nu^2(u)}{\mu^3(u)}
    \frac{d}{du} \nu(u)
    - \frac{ \nu(u) [\nu^2(u) + \mathcal{K}] } {\mu^4(u)} \frac{d}{du} \mu(u) \right\}
    du +
    \\ & & \frac{1}{\mathcal{K}} \left\{ \frac{1}{3} \mu^3(t)
    \ell^3 + \mu(t) \nu(t) \ell^2 + \left[ \frac{\nu^2(t) +\mathcal{K}}{\mu(t)}
    \right] \ell \right\}.\label{T soln flat}
\end{eqnarray}
Recall that in these expression, $\mu$ and $\nu$ can be regarded
as free functions.  Taken with (\ref{R soln}), these equations
give the transformation from \textsc{tbh} to \textsc{lmw}
coordinates explicitly.

Before moving on, there is one special case that we want to
highlight.  This is defined by $k\mathcal{K} < 0$, which implies
that there is no Killing horizon in the bulk for real values of
$R$ and we have a naked singularity.  If we have a spherical
3-geometry, then this is the case of a negative mass black hole.
We have that $\sqrt{k\mathcal{K}} = i \sqrt{-k\mathcal{K}}$, which
allows us to rewrite equation (\ref{T soln curved}) as
\begin{eqnarray}\nonumber
    {\mathcal T}(t,\ell) & = & \frac{1}{k} \left\{ \mu(t) \ell +
    \frac{\mathcal{K}}{\sqrt{-k\mathcal{K}}} \, \mathrm{arctan}
    \left( \frac{k}{\sqrt{-k\mathcal{K}}}
    \frac{[\mu^2(t) +k]\ell + \nu(t)}{\mu(t)} \right) \right\} +
    \\ & & \frac{1}{k} \int_t \left\{ \frac{1}{\mu(u)} \frac{d}{du}\nu(u)-
    \left[ \frac{\nu(u)}{\mu^2(u) + k} \right] \frac{d}{du} \mu(u) \right\}
    du .
\end{eqnarray}
In obtaining this, we have made use of the identity
\begin{equation}\label{identity}
    \mathrm{arctan}\,z = \frac{1}{2i} \ln \frac{1 + iz}{1 - iz}, \quad z \in \mathbb{C}.
\end{equation}

To summarize this section, we have successfully found a coordinate
transformation between the \textsc{tbh} to \textsc{lmw}
coordinates.  This establishes that those two solutions are indeed
isometric, and are hence equivalent.

\subsection[Schwarzschild to Fukui-Seahra-Wesson
coordinates]{Transformation from Schwarzschild to
Fukui-Seahra-Wesson coordinates} \label{sec:coord FSW}

We now turn our attention to finding a transformation between the
\textsc{tbh} and \textsc{fsw} line elements.  The procedure is
very similar to the one presented in the previous section.  We
begin by applying the following general coordinate transformation
to the \textsc{tbh} solution (\ref{BH metric}):
\begin{equation}
    T = {\mathsf T}(\tau,w), \quad R = {\mathsf
    R}(\tau,w).
\end{equation}
Again, instead of identifying $\mathsf{R}(\tau,w) = b(\tau,w)$ as
given by (\ref{FSW scale}), we regard it as a function to be
solved for. To match the metric resulting from this transformation
with (\ref{FSW metric}) we demand
\begin{subequations}\label{FSW transformation}
\begin{eqnarray}\label{one prime}
    +1 & = & h({\mathsf R}) \mathsf{T}_{,\tau}^2 - \frac{ \mathsf{R}_{,t}^2 }{
    h({\mathsf R}) }, \\ \label{two prime} 0 & = & h({\mathsf R})
    \mathsf{T}_{,\tau} \mathsf{T}_{,w}  -
    \frac{ \mathsf{R}_{,\tau} \mathsf{R}_{,w}  }{h({\mathsf R})}, \\ \label{three prime} -
    \frac{\mathsf{R}_{,w}^2}{\zeta^2(w)}  & = & h({\mathsf R})
    \mathsf{T}_{,w}^2 - \frac{ \mathsf{R}_{,w}^2 }{
    h({\mathsf R}) }.
\end{eqnarray}
\end{subequations}
Here, $\zeta(w)$ is an arbitrary function.  Compare this to the
previous system of \textsc{pde}s (\ref{the transformation}).  We
have essentially swapped and changed the signs of the lefthand
sides of (\ref{one}) and (\ref{three}), as well as replaced
$\mathcal{R}_{,t}$ with $\mathsf{R}_{,w}$ and $\mu(t)$ with
$\zeta(w)$.  This constitutes a sort of identity exchange $t
\rightarrow w$ and $\ell \rightarrow \tau$.  The explicit form of
the \textsc{tbh} metric after this transformation is applied is
\begin{equation}
    ds^2_\text{\textsc{tbh}} = d\tau^2 - {\mathsf R}^2(\tau,w) \, d\sigma_{(k,3)}^2 - \left[ \frac{
    \mathsf{R}_{,w}(t,w)}{ \zeta(w)} \right]^2 dw^2.
\end{equation}
This matches the \textsc{fsw} metric \ansatz~(\ref{FSW metric}),
but the functional form of $\mathsf{R}(\tau,w)$ is yet to be
determined by the coordinate transformation (\ref{FSW
transformation}).

Let us now determine it by repeating the manipulations of the last
section.  We find that $\mathsf{R}$ satisfies the \textsc{pde}
\begin{equation}
    \mathsf{R}_{,\tau} = \pm \sqrt{ \zeta^2(w) -
    h({\mathsf R}) },
\end{equation}
which is solved by
\begin{equation}\label{R solution}
    {\mathsf R}^2(\tau,w) = [\zeta^2(w) - k] \tau^2 + 2
    \chi(w) \tau + \frac{ \chi^2(w) - \mathcal{K} }{
    \zeta^2(w) - k}.
\end{equation}
Here, $\chi$ is an arbitrary function.  In a manner similar to
before, we see that the coordinate transformation fixes the
solution for $\mathsf{R}(\tau,w)$, and that it matches the
solution for $b(\tau,w)$ obtained directly from the 5-dimensional
vacuum field equations (\ref{FSW scale}).

The solution for ${\mathsf T}$ is obtained without difficultly as
before. For $k = \pm 1$, we get
\begin{subequations}\label{T solution curved}
\begin{eqnarray}\nonumber
    {\mathsf T}(\tau,w) & = & \frac{1}{k} \int_w \left\{ \frac{1}{\zeta(u)}
    \frac{d}{du}\chi(u)- \left[ \frac{\chi(u)}{\zeta^2(u) - k} \right] \frac{d}{du}
    \zeta(u) \right\} du +
    \\ & & \frac{1}{k} \left\{ \zeta(w) \tau -
    \frac{\mathcal{K}}{2\sqrt{ k\mathcal{K}}} \, \ln
    \frac{1+\mathsf{X}(\tau,w) }{1-\mathsf{X}(\tau,w) } \right\},
    \\ \mathsf{X}(\tau,w) & \equiv & \frac{k}{\sqrt{k\mathcal{K}}}
    \frac{[\zeta^2(w) - k]\tau + \chi(w)}{\zeta(w)}.
\end{eqnarray}
\end{subequations}
For $k = 0$, we obtain
\begin{eqnarray}\nonumber
    {\mathsf T}(\tau,w) & = & \frac{1}{\mathcal{K}} \int_w  \left\{
    \frac{\chi^2(u)}{\zeta^3(u)} \frac{d}{du} \chi(u)
    - \frac{ \chi(u) [\chi^2(u) - \mathcal{K}] } {\zeta^4(u)} \frac{d}{du}
    \zeta(u) \right\} du + \\ & & \frac{1}{\mathcal{K}} \left\{ \frac{1}{3} \zeta^3(w)
    \tau^3 + \zeta(w) \chi(w) \tau^2 + \left[ \frac{\chi^2(w) - \mathcal{K}}{\zeta(w)}
    \right] \tau \right\}. \label{T solution flat}
\end{eqnarray}
These transformations (equations \ref{R solution}--\ref{T solution
flat}) are extremely similar to the ones derived in the previous
section.  Just as before, there are special issues with the
$k\mathcal{K} < 0$ case that can be dealt with using the identity
(\ref{identity}); but we defer such a discussion as it does not
add much to what we have established.

In conclusion, we have succeeded in finding a coordinate
transformation from the \textsc{tbh} to \textsc{fsw} metrics.
Since we have already found a transformation from \textsc{tbh} to
\textsc{lmw}, this allows us to also conclude that a coordinate
transformation between the \textsc{fsw} and \textsc{lmw} metrics
exists as well.

\subsection{Comment: the generalized Birkhoff
theorem}\label{sec:birkhoff}

Before moving on, we would like to make a comment about how the
equivalence of the \textsc{lmw}, \textsc{fsw}, and \textsc{tbh}
metrics relates to the issue of a generalized version of the
Birkhoff theorem. Both the \textsc{lmw} and \textsc{fsw}
\emph{ansatzs} are of the general form:
\begin{equation}\label{general ansatz}
    ds^2 = A^2(t,\ell) \, dt^2 - B^2(t,\ell) \, d\sigma^2_{(k,3)}
    - C^2(t,\ell) \, d\ell^2.
\end{equation}
To this line element, we can apply the coordinate transformation
\begin{equation}
    R = B(t,\ell)
\end{equation}
to obtain
\begin{equation}
    ds^2 = P^2(t,R) \, dt^2 - R^2 \, d\sigma^2_{(k,3)}
    - 2 N(t,R) \, dt\,dR - Q^2(t,R) \, dR^2.
\end{equation}
Here, $P$, $Q$ and $N$ are related to the original metric
functions $A$, $B$, and $C$, but their precise form is irrelevant.
Then, we apply the diffeomorphism
\begin{equation}
    dt = M(T,R) \, dT + \frac{N(t,R)}{P^2(t,R)} \, dR,
\end{equation}
where $M(T,R)$ is an integrating factor that should satisfy
\begin{equation}
    \frac{1}{M} \frac{ \di M}{\di R} = \frac{\di}{\di t}
    \frac{N}{P^2},
\end{equation}
in order to ensure that $dt$ is a perfect differential.  In these
coordinates, the line element is
\begin{equation}\label{line element}
    ds^2 = f(T,R) \, dT^2 - g(T,R) \, dR^2  - R^2 \, d\sigma^2_{(k,3)}.
\end{equation}
Again, $f$ and $g$ are determined by the original metric functions
and the integrating factor.  This structure is strongly
reminiscent of the general spherically-symmetric metric from
4-dimensional relativity.  The only difference is that the line
element on a unit 2-sphere $d\Omega^2$ has been replaced by
$d\sigma^2_{(k,3)}$.  In the 4-dimensional case, Birkhoff's
theorem tells us that the only solution to the vacuum field
equations with the general spherically-symmetric line element is
the Schwarzschild metric.  The theorem has been extended to the
multi-dimensional case by Bronnikov \& Melnikov \cite{Bro95}, who
showed that the 5-dimensional vacuum solution with line element
(\ref{line element}) is unique and given by the \textsc{tbh}
metric.  So, in retrospect it is perhaps apparent that the
\textsc{lmw}, \textsc{fsw}, and \textsc{tbh} solutions are
equivalent --- any 5-dimensional vacuum solution that can be cast
in the form of (\ref{general ansatz}) must be isometric to the
\textsc{tbh} metric.  We conclude by noting that this type of
argument extends to the case of 5-dimensional Einstein spaces as
well, because another variation of Birkhoff's theorem derived by
Bronnikov \& Melnikov is applicable. That is, if there is a
cosmological constant in the bulk --- as in the popular Randall \&
Sundrum braneworld models --- a metric solution of the form
(\ref{general ansatz}) will be equivalent to a deSitter or
anti-deSitter \textsc{tbh} manifold.  For example, the
``wave-like'' solutions sourced by a cosmological constant found
by Ponce de Leon \cite{Pon02} should be isometric to 5-dimensional
Schwarzschild-AdS black holes.

\section[Penrose-Carter embedding diagrams]
{Penrose-Carter diagrams of \textsc{flrw} models embedded in the
Liu-Mashhoon-Wesson metric} \label{sec:coord Kruskal}

We have now established that the \textsc{lmw}, \textsc{fsw}, and
\textsc{tbh} solutions of the vacuum field equations are mutually
isometric.  This means that they each correspond to coordinate
patches on the same 5-dimensional manifold.  Now, it is well-known
that the familiar Schwarzschild solution in four dimensions only
covers a portion of what is known as the extended Schwarzschild
manifold \cite{Kru60}. It stands to reason that if there is a
Killing horizon in the \textsc{tbh} metric, then the $(T,R)$
coordinates will also only cover part of some extended manifold
$M$. This raises the question: what portion of the extended
manifold $M$ is covered by the $(t,\ell)$ or $(\tau,w)$
coordinates?  This is interesting because it is directly related
to the issue of what portion of $M$ is spanned by the universes
embedded on the $\Sigma_\ell$ and $\Sigma_w$ hypersurfaces.

We do not propose to answer these questions for all possible
situations because there are a wide variety of choices of free
parameters.  We will instead concentrate on one particular
problem: namely, the manner in which the Liu-Mashhoon-Wesson
coordinates cover the extended manifold $M$ when $k = +1$,
$\mathcal{K} > 0$, and for specific choices of $\mu$ and $\nu$.
The restriction to spherical $S_3$ submanifolds means that the
maximal extension of the $(T,R)$ coordinate patch proceeds
analogously to the 4-dimensional Kruskal construction. The
calculation can be straightforwardly generalized to the
Fukui-Seahra-Wesson coordinates if desired.

\begin{figure}
\begin{center}
\includegraphics{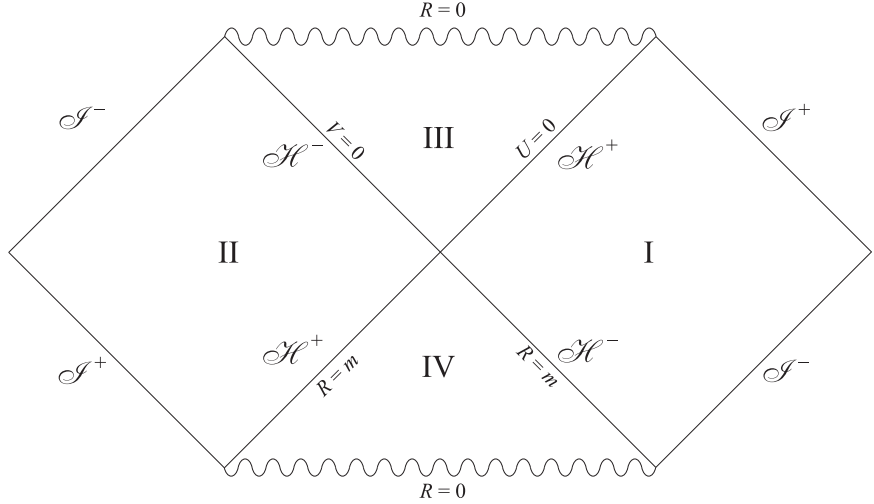}
\caption[Penrose-Carter diagram of a 5D black hole
manifold]{Penrose-Carter diagram of a 5-dimensional black hole
manifold}\label{fig:kruskal}
\end{center}
\end{figure}

We first need to find the 5-dimensional generalization of
Kruskal-Szekeres coordinates for the $k = +1$ \textsc{tbh} metric.
(See ref.~\onlinecite{Mis70} for background information about the
4-dimensional formalism.)  Let us apply the following
transformations to the metric (\ref{BH metric}):
\begin{equation}
    R_* = R + \frac{1}{2} m \ln \left| \frac{R -
    m}{R + m} \right|, \quad u = T - R_*, \quad
    v = T + R_*,
\end{equation}
where we have defined $\mathcal{K} \equiv m^2$ such that the event
horizon is at $R = m$. We then obtain
\begin{equation}
    ds^2_\text{\tiny{BH}} = \text{sgn} \, h(R) \, \frac{(R^2 +
    m^2)e^{-2R/m}}{R^2} e^{-u/m}
    e^{v/m} \, du \, dv - R^2 \,d\Omega_3^2.
\end{equation}
Here, we have changed the ``\textsc{tbh}'' label to
``\textsc{bh}'' to stress that we are dealing with an ordinary
black hole with spherical symmetry.  This metric is
singularity-free at $R = m$. The next transformation is given by
\begin{equation}
    \tilde{U} = \mp \text{sgn} \, h(R) \, e^{-u/m}, \quad \tilde{V} = \pm
    e^{v/m},
\end{equation}
which puts the metric in the form
\begin{equation}
    ds^2_\text{\tiny{BH}} = m^2 \left( 1+ \frac{m^2}{R^2} \right) e^{-2R/m}
    \, d\tilde{U} d\tilde{V} - R^2 \,d\Omega_3^2.
\end{equation}
This is very similar to the 4-dimensional Kruskal-Szekeres
coordinate patch on the Schwarzschild manifold.  The aggregate
coordinate transformation from $(T,R)$ to $(\tilde{U},\tilde{V})$
is given by
\begin{subequations}\label{final transformation}
\begin{eqnarray}
    \tilde{U} & = & \mp \text{sgn} \, h(R) \, e^{-T/m} e^{R/m} \sqrt{\left| \frac{R -
    m}{R + m} \right|}, \\
    \tilde{V} & = & \pm e^{T/m} e^{R/m} \sqrt{\left| \frac{R -
    m}{R + m} \right|}.
\end{eqnarray}
\end{subequations}
From these, it is easy to see that the horizon corresponds to
$\tilde U \tilde V = 0$.  Now, what are we to make of the sign
ambiguity in these coordinate transformations? Recall that in four
dimensions, the extended Schwarzschild manifold involves two
copies of the ordinary Schwarzschild spacetime interior and
exterior to the horizon. It is clear that something analogous is
happening here: the mapping $(T,R) \rightarrow (\tilde U, \tilde
V)$ is double-valued because the original $(T,R)$ coordinates can
correspond to one of two different parts of the extended manifold.
This is best illustrated with a Penrose-Carter diagram, which is
given in Figure \ref{fig:kruskal}.  As is the usual practice, to
obtain such a diagram we ``compactify'' the $(\tilde U, \tilde V)$
coordinates by introducing
\begin{equation}
    U = \frac{2}{\pi} \text{arctan} \, \tilde{U}, \quad
    V = \frac{2}{\pi} \text{arctan} \, \tilde{V}.
\end{equation}
Figure \ref{fig:kruskal} has all of the usual properties: null
geodesics travel on $45^\circ$ lines, the horizons appear at $U =
0$ or $V=0$, the singularities show up as horizontal features at
the top and bottom, and each point in the two-dimensional plot
represents a 3-sphere.  Also, in quadrant I the $T$-coordinate
increases from bottom to top, while the reverse is true in
quadrant II.  We see that the top sign in the coordinate
transformation (\ref{final transformation}) maps $(T,R)$ into
regions I or III of the extended manifold where $V > 0$, while the
lower sign defines a mapping into II or IV where $V < 0$.

\subfiguresbegin\label{fig:embedding}
\begin{figure}[t]
\begin{center}
\includegraphics{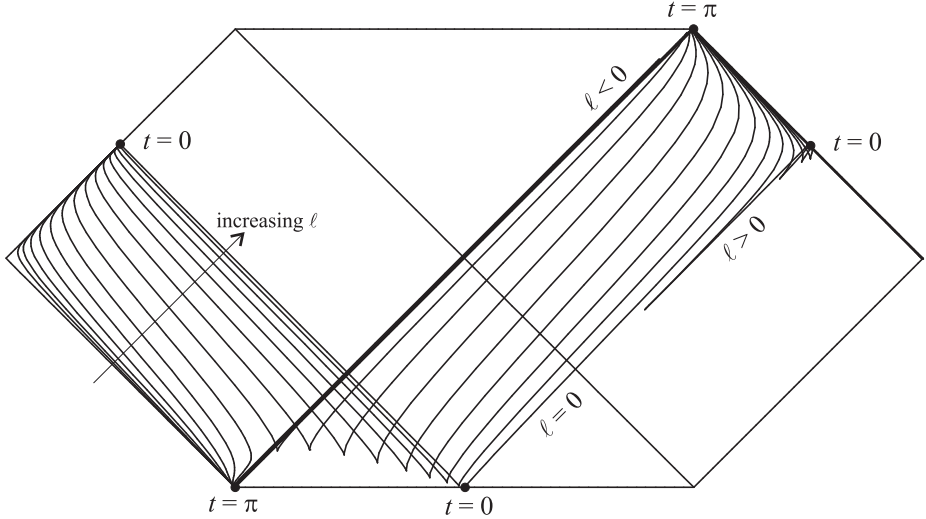}
\caption[$\Sigma_\ell$ hypersurfaces of the \textsc{lmw}
metric]{$\Sigma_\ell$ hypersurfaces of the \textsc{lmw} metric for
the special choices (\ref{choices}).  Each point in the
Penrose-Carter diagram represents a 3-sphere.  We restrict $t \in
(0,\pi)$.  The corresponding values of $\ell$ range from $\sim
-2.2$ to $0.3$ in equal logarithmic intervals.  Note that even
though the two points marked $t=\pi$ appear to be on the $U=0$
line, they are actually located on $\mathcal{I}^+$ in region II
and $R=0$ in region III.  This can be explicitly confirmed by
greatly enlarging the scale of the plot.}\label{fig:isoell}
\end{center}
\end{figure}
\begin{figure}[t]
\begin{center}
\includegraphics{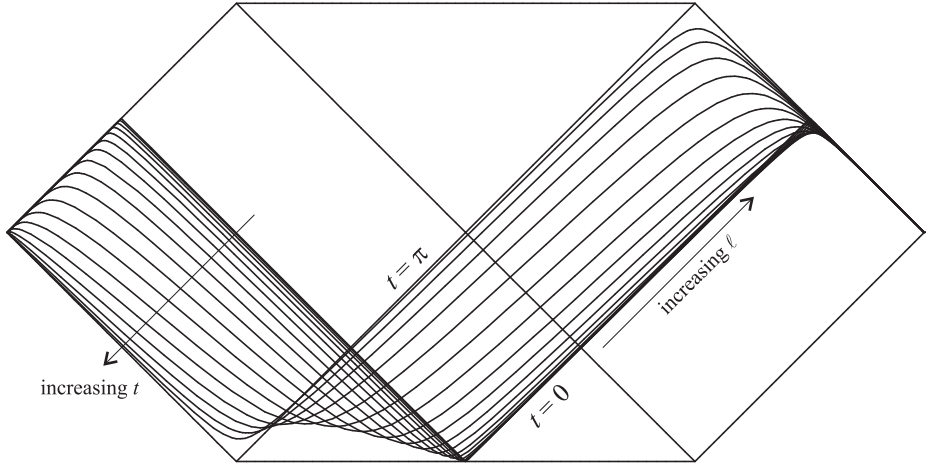}
\caption[Isochrones of the \textsc{lmw} metric]{Isochrones of the
\textsc{lmw} metric for the special choices (\ref{choices}).  We
restrict $\ell \in (-5,5)$.  The corresponding values of $t$ range
from $0$ to $\sim \pi/2$ in equal logarithmic intervals. A portion
of the $t = \pi$ surface is also shown, which appears to be
coincident with $\mathcal{H}^+$.  However, in reality it is only
parallel to $U=0$, but the finite separation between the two
surfaces is impossible to discern without greatly enlarging the
scale of the plot.}\label{fig:isochrone}
\end{center}
\end{figure}
\subfiguresend

Having obtained the transformation to Kruskal-Szekeres
coordinates, we can now plot the trajectory of the $\Sigma_\ell$
hypersurfaces through the extended manifold by using (\ref{R
soln}) and (\ref{T soln curved}) in (\ref{final transformation})
to find $U(t,\ell)$ and $V(t,\ell)$.  But there is one wrinkle: we
need to flip the sign of the $(T,R) \rightarrow (U,V)$
transformation whenever the path crosses the $V=0$ line, which is
not hard to accomplish numerically.  In Figure
\ref{fig:embedding}, we present Penrose-Carter embedding diagrams
of $\Sigma_\ell$ and $\Sigma_t$ hypersurfaces associated with the
\textsc{lmw} metric for the following choices of parameters and
free functions:
\begin{equation}\label{choices}
    m = \tfrac{1}{2}, \quad
    \mu(t) = \cot t, \quad
    \nu(t) = \tfrac{\sqrt{3}}{2}.
\end{equation}
This gives
\begin{equation}
    a(t,\ell) = \sqrt{\left( \ell \csc t + \tfrac{\sqrt{3}}{2} \sin t
    \right)^2 + \tfrac{1}{4} \sin^2 t}.
\end{equation}
Our choices imply that it is sensible to restrict $t \in (0,\pi)$.
For $\ell \ne 0$, the cosmologies embedded on $\Sigma_\ell$ do
\emph{not} undergo a big bang or big crunch and $a \rightarrow
\infty$ as $t \rightarrow 0$ or $\pi$. The $\ell = 0$ cosmology
simply has $a(t,0) = \sin t$.  That is, we have a re-collapsing
model. The induced metric for that hypersurface is
\begin{equation}
    \ds{\Sigma_0} = \sin^2 t \, (dt^2 - d\Omega_3^2),
\end{equation}
that of a closed radiation-dominated universe.

In Figure \ref{fig:isoell} we show the $\Sigma_\ell$ hypersurfaces
of this model in a Penrose-Carter diagram.  In this plot it is
easy to visually determine where each trajectory begins when $t
=0$, but because of the scale it is difficult to note precisely
where they end up at when $t = \pi$.  By careful analysis of the
numeric results, we have determined the following facts:  The
$\ell = 0$ trajectory emanates from the middle of the singularity
in the white hole region IV at $t = 0$ and terminates on the
future singularity in region III at $t = \pi$.  The surfaces with
$\ell > 0$ begin at $\mathcal{I}^+$ in I and terminate on
$\mathcal{I}^-$ in the same region.  The models with $\ell < 0$
all begin on $\mathcal{I}^-$ and terminate on $\mathcal{I}^+$ in
region II.  We mention in passing that this plot bears some
qualitative resemblance to the figures of Mukohyama et
al.~\cite{Muk99}, who showed the equivalence of a known solution
of the 5-dimensional field equations with a cosmological constant
and the topological Schwarzschild-AdS black hole in the context of
braneworld scenarios; but many details are significantly
different.

One of the most striking features of this plot is the cusps
present in the majority of the $\Sigma_\ell$ curves.   These sharp
corners suggest some sort of singularity in the embedding at their
location.  We can search for the singularity by examining scalars
formed from the extrinsic curvature of the $\Sigma_\ell$
4-surfaces.  Let us consider
\begin{equation}
    h^{\alpha\beta} K_{\alpha\beta} =  \frac{a_{,t\ell}}{a_{,t}} +
    3 \frac{a_{,\ell}}{a}.
\end{equation}
One can confirm directly that this diverges whenever $a_{,t} = 0$
and $a_{,t\ell} \ne 0$.  At such positions, we find sharp corners
in the $\Sigma_\ell$ hypersurfaces.  This makes it clear that if
we wanted to use the \textsc{lmw} coordinates as a patch on the
extended 5-dimensional black hole manifold, we would have to
restrict $t$ to lie in an interval bounded by times defined by the
turning points of $a$.  This is in total concurrence with the
analysis of singularities in the intrinsic 4-geometry performed in
Section \ref{sec:LMW} --- the cusps correspond to singularities in
the induced metric on $\Sigma_\ell$.  Actually we have confirmed
that the curves with cusps generally have two curvature anomalies,
but those additional features tend to get compressed into a region
too small to resolve in Figure \ref{fig:isoell}.  What is also
interesting about these plots is how the \textsc{lmw} metric
occupies a fair bit of territory in $M$ (some of the $\Sigma_\ell$
hypersurfaces span regions I, II and IV). Like the
Kruskal-Szekeres coordinates, the \textsc{lmw} patch is regular
across the horizon(s).

The exact portion of the extended manifold spanned by our model is
a little clearer in Figure \ref{fig:isochrone}.  In this plot, we
show the $\Sigma_t$ spacelike hypersurfaces --- or isochrones
--- of the \textsc{lmw} metric.  These stretch from spacelike infinity in
region II to a point on $\mathcal{I}^+$ in region I.  The
\textsc{lmw} time $t$ is seen to run from bottom to top in I and
\emph{vice versa} in II.  We also see clearly that there is a
portion of the white hole region IV that is not covered by the
\textsc{lmw} metric with $t \in (0,\pi)$.  The $t = \pi$ line
appears to coincide with $U = 0$, but is in actuality displaced
slightly to the left.  Notice that the area bounded by the
$t=\pi/2$ and $t = \pi$ curves is relatively small, from which it
follows that the portions of the $\Sigma_\ell$ surfaces with
$\pi/2 \lesssim t \lesssim \pi$ tend to occupy an extremely
compressed portion of the embedding diagram.

In summary, we have presented embedding diagrams for the
$\Sigma_\ell$ and $\Sigma_t$ hypersurfaces associated with the
\textsc{lmw} metric in the Penrose-Carter graphical representation
of the extended 5-dimensional black hole manifold.  This partially
answers the question of which portion of $M$ is occupied by the
\textsc{lmw} metric. However, the calculation was for specific
choices of $\mu$, $\nu$, and $\mathcal{K}$.  We have no doubt that
more general conclusions are attainable, but that is a subject for
a different venue.

\section{Summary and Discussion}\label{sec:summary}

In this paper, we introduced two solutions of the 5-dimensional
vacuum field equations, the Liu-Mashhoon-Wesson and
Fukui-Seahra-Wesson metrics, in Sections \ref{sec:LMW} and
\ref{sec:FSW} respectively. We showed how both of these embed
certain types of \textsc{flrw} models and studied the coordinate
invariant properties of the associated 5-manifolds.  We found that
both solutions had line-like curvature singularities and Killing
horizons, and that their Kretschmann scalars were virtually
identical.  These coincidences prompted us to suspect that the
\textsc{lmw} and \textsc{fsw} metrics are actually equivalent, and
that they are also isometric to the 5-dimensional topological
black hole metric introduced in Section \ref{sec:topological}.
This was confirmed explicitly in Section
\ref{sec:transformations}, where transformations from
Schwarzschild-like to \textsc{lmw} and \textsc{fsw} coordinates
were derived.  The strategy employed in that section was to
transform the \textsc{tbh} line element into the form of the
\textsc{lmw} and \textsc{fsw} metric \emph{ansatzs}, which
resulted in two sets of solvable \textsc{pde}s. Therefore, those
calculations comprise independent derivations of the \textsc{lmw}
and \textsc{fsw} metrics.  In Section \ref{sec:birkhoff}, we
showed how the relationship between the \textsc{lmw}, \textsc{fsw}
and \textsc{tbh} metrics was a consequence of a generalized
version of Birkhoff's theorem. Finally, in Section \ref{sec:coord
Kruskal} we performed a Kruskal extension of the 5-dimensional
black hole manifold and plotted the $\Sigma_\ell$ and $\Sigma_t$
hypersurfaces of the \textsc{lmw} metric in a Penrose-Carter
diagram for certain choices of $\mu$, $\nu$, and $\mathcal{K}$.

Obviously, our main result is that the \textsc{lmw} and
\textsc{fsw} metrics are non-trivial coordinate patches on
5-dimensional black hole manifolds.  We saw explicitly that the
\textsc{lmw} coordinates could cover multiple quadrants of the
maximally-symmetric manifold, and that they were regular across
the event horizon.  This puts them in the same category as the
Eddington-Finkelstein (\textsc{ef}) or Painlev\'{e}-Gullstrand
(\textsc{pg}) coordinates associated with 4-dimensional
Schwarzschild black holes \cite{Mar00}, which are also horizon
piercing patches that do not involve implicit functions, such as
$R = R(U,V)$ in the Kruskal-Szekeres covering. The \textsc{lmw}
coordinates differ from the \textsc{ef} or \textsc{pg} patches in
that they are 5-dimensional and orthogonal. All of these features
make them an attractive tool for the study of black hole physics
in 5 dimensions.  In particular, they provide ``rest-frame''
coordinates for embedded 4-dimensional universes. That is, in both
the \textsc{lmw} and \textsc{fsw} coordinates, universes are
defined simply as 4-surfaces comoving in $\ell$ or $w$.  And
unlike standard Schwarzschild-like coordinates, the \textsc{lmw}
or \textsc{fsw} 5-metrics are regular as the universe crosses the
black hole horizon(s). Such coordinates may have some utility in
the study of quantized braneworld models, where the bad behaviour
of coordinates across horizons apparently results in a complicated
canonical phase-space description of the brane's dynamics
\cite{Sea03a}.

Finally, we discuss the temptation to generalize these coordinates
to other types of black holes and different dimensions.  One could
easily imagine repeating the manipulations of Section
\ref{sec:transformations} for different choices of $h(R)$, which
could be selected to correspond to any spherically-symmetric black
hole in any dimension.  However, a difficulty arises when one
tries to integrate equations like
\begin{equation}
    {\mathcal R}_{,\ell} = \pm \sqrt{h({\mathcal R}) + \mu^2(t)}.
\end{equation}
to obtain $\mathcal{R} = \mathcal{R}(t,\ell)$ explicitly.  It
turns out that this in not necessarily easy to do if $h({\mathcal
R}) \ne k - \mathcal{K}/{\mathcal R}^2$.  For example, if $h$
corresponds to an $N$-dimensional topological black hole (i.e.,
$h({\mathcal R}) = k - \mathcal{K}/{\mathcal R}^{N-3}$) we obtain
complicated implicit definitions of $\mathcal{R}$ involving
generalized hypergeometric functions.  For even $N = 4$, it is
unclear how to invert such an equation to find $\mathcal{R} =
\mathcal{R}(t,\ell)$ explicitly.  So it seems that the
5-dimensional case is somewhat special.  However, we do not
preclude the possibility that there are other special cases out
there, that our procedure could be improved upon, or that one
could find suitable coordinates by direct assault on the
$N$-dimensional field equations.  Such issues are best addressed
by future work.

\begin{acknowledgments}
We would like to thank T.~Liko for useful discussions, and NSERC
and OGS for financial support.
\end{acknowledgments}

\bibliography{black_hole}

\end{document}